# Six Levels of Autonomous Process Execution Management (APEM)

*Wil van der Aalst, Lehrstuhl für Informatik 9 / Process and Data Science, RWTH Aachen University, D-52056 Aachen, Germany, www.vdaalst.com.*

Terms such as the Digital Twin of an Organization (DTO) and Hyperautomation (HA) illustrate the desire to autonomously manage and orchestrate processes, just like we aim for autonomously driving cars. Autonomous driving and *Autonomous Process Execution Management* (APEM) have in common that the goals are pretty straightforward and that each year progress is made, but fully autonomous driving and fully autonomous process execution are more a dream than a reality. For cars, the *Society of Automotive Engineers* (SAE) identified six levels (0-5), ranging from no driving automation (SAE, Level 0) to full driving automation (SAE, Level 5) [1]. This short article defines *six levels* of Autonomous Process Execution Management (APEM). The goal is to show that the transition from one level to the next will be gradual, just like for self-driving cars.

## Relating Autonomous Driving and Autonomous Process Execution Management

The first experiments with self-driving cars were already conducted in the 1930-ties. *Futurama*, a prototype installation sponsored by General Motors, was presented at the 1939 New York World's Fair. Futurama represented a futuristic view of the future of transport using radio-controlled electric cars propelled via electromagnetic fields provided by circuits embedded in the roads. Over time, technology developed at a breathtaking pace, but fully autonomous driving is still a dream and not a reality.

Mercedes-Benz was the world's first automaker to gain international approval to produce a car that is capable of Level 3 autonomous driving in 2022 [3,4]. In 2021, Honda produced a limited set of cars allowed to drive at Level 3 in Japan. However, the usage is limited. E.g., the *Mercedes S-Class with Drive Pilot can only be used at daytime, on highways, and at speeds below 60 kilometers per hour*. It is still not allowed to take a nap while driving, and it seems that real self-driving cars are far from reality. The same applies to autonomous process execution. Since Gartner introduced terms such as the Digital Twin of an Organization (DTO) and Hyperautomation (HA), people have started to think about the use of Machine Learning (ML) and Artificial Intelligence (AI) beyond the automation of individual tasks. Moreover, process mining emerged as a new ingredient supporting such a development. However, buzzwords like DTO and HA are even more confusing than the six levels defined by the Society of Automotive Engineers (SAE) [2]. Therefore, we start by providing a few existing definitions.

## Towards a Digital Twin of an Organization

Gartner uses the following definition for a DTO: "A Digital Twin of an Organization (DTO) is a dynamic software model of any organization that relies on operational and/or other data to understand how an organization operationalizes its business model, connects with its current state, responds to changes, deploys resources and delivers exceptional customer value" [5]. This can be seen as one of the grand challenges in information systems, just like autonomous driving in mobility. Gartner also introduced the related term "hyperautomation" which aims at creating a common set of concepts and technologies to orchestrate all islands of automation, new and existing and at all levels, within an organization [6]. At the same time, we see the uptake of low-code automation frameworks, Robotic Process Automation (RPA), and Task Automation (TA). These novel automation approaches aim to replace human workers by software in a cost-effective manner [8,10]. As a side-effect, processes can be improved, but the main focus is on automation. There is often a vague reference to AI and ML, but the connection is other rather weak, and solutions are often static and programmed or configured by humans.

As defined in the book "Process Mining: Data Science in Action" [9], process mining is designed to discover, monitor, and improve real processes (not assumed processes) by extracting knowledge from event logs readily available in today's information systems. Process mining includes automated process discovery (i.e., extracting process models from an event log), conformance checking (i.e., monitoring deviations by comparing model and log), social network/organizational mining, automated construction of simulation models, model extension, model repair, case prediction, and history-based recommendations. In recent years, the scope of process mining was extended to also include ML, AI, and automation, e.g., action-oriented and predictive process mining. Process mining is clearly one of the building blocks towards APEM.

Process Execution Management (PEM) is close to traditional Business Process Management (BPM). Both PEM and BPM do not focus on the execution of individual tasks, but on the orchestration and management of processes. However, unlike a classical BPM system, a Process Execution Management System (PEMS) does not assume explicitly programmed workflows and is more data-driven. The PEMS is often implemented as a layer on top of existing software systems and people. The Celonis EMS can be seen as a PEMS [7]. Moreover, several other vendors are combining process mining with automation to realize a PEMS.

## Six Levels of Autonomous Process Execution Management

Just like autonomous driving, autonomous PEM is not easy to define. However, it is clear that there are different levels. The table on the next page compares the six SAE levels for autonomous driving [1] with six possible levels of autonomous PEM.

|         | SAE levels for autonomous driving | Levels of autonomous process execution management |
|---------|-----------------------------------|--------------------------------------------------|
| **Level 0** | A human is driving, and features are limited to breaking assistance, blind-spot warning, lane departure warning, etc. | There is no PEMS. All orchestration and management are done by humans. Features are limited to dashboards, reporting, key performance indicators, hard-coded workflows, and manually created simulations to conduct what-if analysis. |
| **Level 1** | A human is driving, but the car provides steering or brake/acceleration support, e.g., lane centering or adaptive cruise control. | The PEMS is able to detect and quantify known and unknown performance and compliance problems. Features include process discovery and conformance checking. The PEMS may create alerts. However, humans need to interpret the diagnostics and, if needed, select appropriate actions. |
| **Level 2** | A human is driving, but the car provides steering and brake/acceleration support. The difference with Level 1 is the combination of systems. | The PEMS is able to detect and quantify known and unknown performance and compliance problems. Moreover, the PEMS is able to recommend actions in case of detected known performance and compliance problems (execution gaps) and support the user in triggering corresponding actions. These actions may be automated, but in-the-end a human decides. |
| **Level 3** | Under selected circumstances, the car is driving. However, the driver needs to be alert and ready to take over control at any time. | The PEMS automatically responds to performance and compliance problems by taking appropriate actions. However, this is limited to a subset of problems and humans need to be alert and ready to take over control. |
| **Level 4** | Under selected circumstances, the car is driving. If the conditions are not met, the vehicle stops. The driver does not need to constantly monitor the situation. | The PEMS automatically responds to performance and compliance problems by taking appropriate actions. In principle, all management and orchestration decisions are made by the PEMS. Humans do not need to constantly monitor the PEMS, but the system may decide to call on the help of humans in case of diverging or unexpected behaviors. |
| **Level 5** | The car can drive itself under all circumstances (comparable to a human driver). | The PEMS functions fully autonomous also under diverging or unexpected circumstances. |

The six possible levels of autonomous PEM should be seen as an initial proposal. It is important to understand that the scope is limited to the orchestration and management of operational processes. Classical workflow and BPM systems operate at Level 0. As of Level 3, the PEMS automatically selects actions based on the current and historical data.

## Hybrid Intelligence

As mentioned before, the Mercedes S-Class can operate at Level 3, but this is limited to highways during daytime and for speeds below 60 kilometers per hour. Just like there are different types of roads and traffic conditions, there are different types of processes operating under different circumstances. Standard processes such as Order-to-Cash (O2C) and Purchase-to-Pay (P2P) are very different from the production of chips or the treatment of Covid patients. Moreover, processes may be in steady-state or experience exceptional loads. Therefore, most levels require an interplay between humans and the PEMS. This matches well with the *hybrid intelligence* concept, which tries to combine human and machine intelligence. The strengths of *human intelligence* are characterized by the words flexible, creative, emphatic, instinctive, and commonsensical. The strengths of *machine intelligence* are characterized by the words fast, efficient, cheap, scalable, and consistent. Hence, *hybrid intelligence will be one of the recurring themes in the context of APEM*.

As Niels Bohr once said, "It is difficult to make predictions, especially about the future" and this applies to both autonomous driving and APEM. In 2015, Elon Musk stated that: "The Tesla that is currently in production has the ability to do automatic steering autopilot on the highway. That is currently being beta tested and will go into a wide release early next month. So, we are probably only a month away from having autonomous driving at least for highways and for relatively simple roads. My guess for when we will have full autonomy is approximately three years." This was clearly too optimistic. The same applies to the claims of software vendors when it comes to AI and ML. However, the direction is clear. The importance of PEM will increase, and PEM systems will become more autonomous over time. However, this will be a slow and gradual process starting with simple processes such as O2C and P2P under "perfect weather conditions".